\documentstyle[12pt]{article}
\begin{document}
\thispagestyle{empty}

\begin{center}
\LARGE \tt \bf{ Has spacetime torsion already been detected in Lorentz and CPT tests with spin-polarized bodies?}
\end{center}

\vspace{2.5cm}

\begin{center} {\large L.C. Garcia de Andrade\footnote{Departamento de
F\'{\i}sica Te\'{o}rica - Instituto de F\'{\i}sica - UERJ

Rua S\~{a}o Fco. Xavier 524, Rio de Janeiro, RJ

Maracan\~{a}, CEP:20550-003 , Brasil.

E-mail : garcia@dft.if.uerj.br}}
\end{center}

\vspace{2.0cm}

\begin{abstract}
Bluhm-Kostelecky (BK) Lagrangian to test Lorentz violation with spin-polarised test bodies is shown to be transformed to a torsion-coupling constant Lagrangian by taking into account the Fock-Ivanenko coefficients in spinor form. In this way one is able to introduce torsion in their Lagrangian and make use of the value of their coupling constant to infer that torsion maybe observed in this type of spin-polarised body experiment.The theory would be a low-energy limit of a quasi-Riemannian gravity in Riemann-Cartan spacetime previously proposed by Gasperini.In this theory of gravity Lorentz local gauge symmetry is violated while global coordinate transformations are preserved.Torsion coupling obtained in this way could be as high as $10^{-29} GeV$. 
\end{abstract}

\vspace{1.0cm}

\begin{center}
\end{center}

\newpage

\pagestyle{myheadings}
\markright{\underline{Torsion and spin-polarised bodies.}}

Several authors \cite{1,2,3,4,5} have proposed alternative methods to test spacetime torsion in nature. In particular from the point of view of high-energy physics Belayev and Shapiro \cite{6} have shown that starting from the Dirac action coupled coupled to the electromagnetic and torsion field there is some additional softly broken symmetry associated with torsion. The requirement of renormalizability in their case fixes the torsion field to some massive pseudovector and its action is fixed with precision to the values of coupling constant of torsion-spinor interaction. In this Letter we show that the torsion-spinor interaction can be obtained in the low-energy form by making use of the Fock-Ivanenko coefficients into the Bluhm-Kostelecky \cite{7} Lagrangian to test CPT and Lorentz symmetries with spin-polarized test bodies. This idea favours Gasperini's \cite{8}  Quasi-Riemannian gravity in Riemann-Cartan spacetime. We argue that BK lagranian would be a low-energy limit Lagrangian of a quasi-Riemannian gravity. This idea is very reasonable since the quasi-Riemannian gravity was introduced in the literature by Weinberg \cite{9} as a gravitational theory which breaks Lorentz local gauge symmetry and preserves the global general coordinate transformation of General Relativity. More recently an investigation of magnetic fields in torsion balance experiments with cylindrically symmetric solutions of Einstein-Cartan-Maxwell field equations was obtained in the literature \cite{3}. Let us consider the BK Lagrangian given by
\begin{equation}
L = -a^{e}_{\mu}J^{\mu} - b^{e}_{\mu} {\bar{\psi}}{\gamma}_{5}{\gamma}^{\mu} {\psi} -\frac{1}{2} H^{e}_{{\mu}{\nu}}{\bar{\psi}}{\sigma}^{{\mu}{\nu}}{\psi} +\frac{1}{2} ic^{e}_{{\mu}{\nu}} {\bar{\psi}}{\gamma}^{\mu} D^{\nu} {\psi}+\frac{1}{2} id^{e}_{{\mu}{\nu}} {\bar{\psi}}{\gamma}_{5}{\gamma}^{\mu}D^{\nu}{\psi}
\label{1}
\end{equation}
where parameters a,b,c and d and H govern the small magnitudes of Lorentz violation,  $iD_{\mu}=i{\partial}_{\mu}-qA_{\mu}$ with charge $q=-e$ and $J^{\mu}={\bar{\psi}}{\gamma}^{\mu}{\psi}$ is the Dirac or Pauli curent depending on the level of energy involved. As it is well-known the torsion tensor can be expressed in spinor form by the Fock-Ivanenko expression \cite{8}
\begin{equation}
S^{{\nu}{\rho}{\mu}}= -\frac{1}{2}{\epsilon}^{{\nu}{\rho}{\mu}{\sigma}}{\bar{\psi}}{\gamma}_{\sigma} {\gamma}_{5}{\psi}
\label{2}
\end{equation}
Substitution of expression (\ref{2}) into expression (\ref{1})
\begin{equation}
L=-a^{e}_{\mu}J^{\mu}-b^{e}_{\mu}S^{\mu}-\frac{1}{2}H^{e}_{{\mu}{\nu}}{\bar{\psi}}{\sigma}^{{\mu}{\nu}}{\psi}+\frac{1}{2}ic^{e}_{{\mu}{\nu}} D^{\nu}J^{\mu}-\frac{1}{2}ic^{e}_{{\mu}{\nu}}D^{\nu}P^{\mu}{\psi}-\frac{1}{2}id^{e}_{{\mu}{\nu}} T^{{\nu}{\mu}}{\psi}-\frac{1}{2}id^{e}_{{\mu}{\nu}}R^{{\nu}{\mu}}{\psi}
\label{3}
\end{equation}
where
\begin{equation}
P^{\mu}={\bar{\psi}}{\gamma}_{5}{\gamma}^{\mu}
\label {4}
\end{equation}
where
\begin{equation}
T^{{\mu}{\nu}}=D^{\nu}S^{\mu} 
\label{5}
\end{equation}
and 
\begin{eqnarray}
R^{{\mu}{\nu}}=D^{\nu}P^{\mu} 
\label{6}
\end{eqnarray}
Here Lagrangian (\ref{3}) is the BK Lagrangian expressed in terms of axionic torsion.This Lagrangian allows us to infer that spacetime torsion has possibly been observed in its low-energy limit form since the coupling constant $b^{e}_{\mu}$ which represents the odd-parameter of Lorentzian violation is also the torsion spinor coupling constant in analogy to the high energy case investigated by Belayev and Shapiro.Besides as computed by BK this coupling is of the order of $10^{-29} GeV$.Just to give an idea of how big is this coupling, recently (\ref{4}) we have shown by considering a conformal breaking Lagrangian for the massive photon in spaces with torsion that an estimation of axial torsion of the order of $10^{-29} eV$ is found.
\section*{Acknowledgement}
I am very much indebt to Prof. G. Gillies and Prof. Ilya Shapiro  for helpful discussions on the subject of this paper. Financial support from CNPq. and UERJ is gratefully acknowledged.

\end{document}